\documentclass[a4paper]{paper} 
\usepackage{hyperref}
\usepackage{amsfonts}
\usepackage{amsmath}
\usepackage{graphicx}
\usepackage[space]{grffile}
\usepackage[margin=2.5cm]{geometry}
\usepackage{pdfpages}
\usepackage[capitalize,noabbrev]{cleveref}

\graphicspath{{img/}}
\title{Imbalance measure and proactive channel rebalancing algorithm for the Lightning Network}
\author{Rene Pickhardt and Mariusz Nowostawski} 
\institution{NTNU Gj{\o}vik} 

\begin{document} 
\twocolumn[\maketitle 
\hrule 
\begin{abstract}
Making a payment in a privacy-aware payment channel network is achieved by trying several payment paths until one succeeds.
With a large network, such as the Lightning Network, a completion of a single payment can take up to several minutes.
We introduce a network imbalance measure and formulate the optimization problem of improving the balance of the network as a sequence of rebalancing operations of the funds within the channels along circular paths within the network.
As the funds and balances of channels are not globally known, we introduce a greedy heuristic with which every node despite the uncertainty can improve its own local balance.
In an empirical simulation on a recent snapshot of the Lightning Network we demonstrate that the imbalance distribution of the network has a Kolmogorov-Smirnoff distance of $0.74$ in comparison to the imbalance distribution after the heuristic is applied.
We further show that the success rate of a single unit payment increases from $11.2\%$ on the imbalanced network to $98.3\%$ in the balanced network.
 Similarly, the median possible payment size across all pairs of participants increases from $0$ to $0.5$ mBTC for initial routing attempts on the cheapest possible path.
We provide an empirical evidence that routing fees should be dropped for proactive rebalancing operations.
Executing $4$ different strategies for selecting rebalancing cycles lead to similar results 
indicating that a collaborative approach within the friend of a friend network might be preferable from a practical point of view.

\end{abstract}

\hrule\bigskip
]

\section{Introduction}
The Lightning Network has been introduced in order to mitigate the scaling issues of blockchain technologies such as Bitcoin\cite{poon2016bitcoin}.
With the help of smart contracts, a payment channel can be established to allow the cryptographically secure transfer of value without the necessity to record the transaction to the central ledger.
This yields enhanced privacy properties for the participants of the network.
In order to protect that privacy while routing a payment through several payment channels privacy aware payment channel networks use a source-based onion routing scheme, like the Sphinx Mix format~\cite{danezis2009sphinx}.
This prevents routing nodes from learning who paid whom.

While the Lightning Network, for example, shares the capacity of public channels with its participants through its gossip protocol the local split of the capacity of the channels into the balance of its participants is not shared with the rest of the network for two reasons:
(1) this would compromise the privacy as one could collect all changes of the channel balances and reconstruct the flow of payments;
(2) propagating this information would essentially mean that every node in the network is made aware of every payment which would have the same poor scaling properties as blockchain technologies and other broadcast networks.
The decision to use source-based routing together with the unknown channel balances of the network results in a challenge for finding a path of payment channels so that all channels along the path have enough liquidity to be able to forward an attempted payment.
Currently, this challenge is met by probing paths with a brute force approach until one such path succeeds to route the payment.
It has been shown that probing for paths can take more than 3 minutes for more than $5\%$ of the attempted payments~\cite{decker2019lnconf}. This leads to a poor payment latency and poor user experience.
Imagine a grocery store in which for every $20^{th}$ customer the cash register would have to wait three minutes until the payment was received.

In this paper we examine the consequences of nodes proactively and collaboratively distributing their funds evenly across their channels.
We introduce the notation of an imbalanced network.
The imbalance is measured as the average of the imbalance scores of its nodes.
The node's imbalance of payment channels is defined as the Gini coefficient of a node's channel balance coefficients which are the relative amount of funds a node owns in a channel in comparison to the capacity of that channel.
Note, nodes on both ends of the channel already have all the information they need to compute their imbalance. 
Nodes can reduce their imbalance either by initiating a rebalancing their own channels themselves, or, by collaboratively rebalancing the channels with the help of their channel partners by conducting circular payments.
Another option is to use a submarine swap\footnote{\url{https://github.com/submarineswaps/swaps-service}} as an off-chain/on-chain swapping service.

We formulate an optimization problem of finding a sequence of rebalancing operations that minimize the network's imbalance.
As the information necessary to solve the optimization problem is not publicly known in privacy-aware payment channel networks, we provide a greedy heuristic for its participants to find a minimum for the problem. 
In an empirical study, we show that the greedy heuristic will lead to a success rate of $98.3\%$ for a small payment between two arbitrary selected nodes.
Also the median possible payment size between all pairs of nodes increases from $0$ in the imbalanced network to $0.5$ mBTC in the best balanced networks that we found, when the cheapest possible path is selected for the payment.

While single nodes can already execute the algorithm in currently existing payment channel networks they have to pay routing fees for the rebalancing operations.
Thus economically, the fees disincentivize the nodes to conduct rebalancing operations.
We compared the earned fees for being participant of another nodes rebalancing attempt and the paid fees for initiating rebalancing operations.
Showing that these numbers are normally distributed around $0$ we propose to omit fees for proactive rebalancing operations to incentivize nodes to execute our algorithm which appears to be beneficial for the entire network. 
Looking at different strategies of finding rebalancing cycles we suggest, for the sake of speed, that nodes share with their neighbors on which local channels they would like to have inbound or outbound capacity.
This information could easily be probed anyway and will not to worsen the privacy of the nodes involved in rebalancing.

The remainder of this paper is organized as follows. In~\cref{sec:relatedWork} we give a short overview of related work in this field.
We then introduce the notation and definition of the imbalance score in~\cref{sec:formalization} and formulate the optimization problem to reduce the imbalance.
We propose a greedy and collaborative algorithm in~\cref{sec:Algorithm} to address the optimization problem.
We also propose different strategies for the algorithm to probe for potential rebalancing cycles.
In~\cref{sec:setup} we introduce the experimental setup and the data set that we used.
After showing our empirical results in~\cref{sec:results} we discuss the results, 
we propose extensions and future work in the last two sections.

\section{Related Work}
\label{sec:relatedWork}

While the Lightning Network white paper~\cite{poon2016bitcoin} does not discuss path finding and states routing as an easy problem it is generally recognized that pathfinding on the Lightning Network is a difficult problem~\cite{piatkivskyi2018split, prihodko2016flare, bagaria2019boomerang, pickhardt2019pathfinding, grunspan2018ant, sivaraman2018routing}.
There is already research conducted in the field of rebalancing channels~\cite{khalil2017revive} which was more about the cryptographic protocols used to make sure that participants can enforce the rebalancing that was agreed upon.
There are rebalancing operations for c-lightning\footnote{\url{https://github.com/lightningd/plugins/tree/master/rebalance}} and for lnd\footnote{\url{https://github.com/bitromortac/lndmanage}}.
In particular, the idea of just in time rebalancing while fulfilling routing requests~\cite{pickhardt2019jit} has been implemented as JIT-routing for c-lightning\footnote{\url{https://github.com/lightningd/plugins/pull/66}}.

\section{Formalization and Assumptions}
\label{sec:formalization}

Let $N=(V,E,c)$ be a payment channel network with a finite set of nodes.
The payment channels are the edges in the network such that $E\subset V\times V$.
Additionally, we have a publicly known capacity function $c: E\longrightarrow \mathbb{N}$ that assigns a capacity to every edge of the network.
For every edge $e=(u,v)$ we denote $e_u:=(e,u)$ as the first participant of the channel and $e_v=(e,v)$ as the second participant.
Naturally, the capacity of every channel $e=(u,v)$ is privately split into the local balances with the balance function $b: E\times V\longrightarrow\mathbb{N}$ such that $b(e_u)+b(e_v)\stackrel{!}{=}c(e)$.

We define the channel balance coefficient for $u$ on the channel $e=(u,v)$ as  $\zeta_{(u,v)} = \frac{b(e_u)}{c(e)}$.
This is just the relative amount of funds that the participant $u$ has in the channel $e$.
Since $b(e_u)+b(e_v)\stackrel{!}{=}c(e)$ we also have $\zeta_{(u,v)} + \zeta_{(v,u)}=1$.
In general, for an imbalanced chanel we have $\zeta_{(u,v)} \neq \zeta_{(v,u)}$.

The neighbor function $n : V \longrightarrow 2^{E}$ assigns every node a set of all the channels it is part of.
To make the following formulas easier to read let us introduce the neighborhood $U:=n(u)$.

The total funds of a participant $u$ are denoted as $\tau_u:=\displaystyle{\sum_{e\in U}b(e_u)}$.
The value of $\tau_u$ is constant while no payments are made and no channels are being opened and closed.
In contrast the balance function $b$ can vary subject to re-allocation of funds.
The total capacity of a participant $u$ is denoted as $\kappa_u:=\displaystyle{\sum_{e\in U}c(e)}$.
Using the last two definitions let us define the node balance coefficient for a participant $u$ as $\nu_u = \frac{\tau_u}{\kappa_u}$.

We call a node $u$ {\bf balanced} if its channel balance coefficients $\zeta_{(u,v_1)},\dots,\zeta_{(u,v_d)}$ have the same value.\footnote{This value does by no means have to take a value of $0.5$ though it certainly seems to be a reasonable goal.}
This means that the distribution of a node's relative funds across all its channels is the same.
Consequently, we consider a node {\bf unbalanced}, if it local channel balance coefficients are unequal.
Statistically, inequality of a distribution can be measured with the Gini coefficient.
Thus, for a node $u$ with channel balance coefficients $\zeta_{(u,v_1)},\dots,\zeta_{(u,v_d)}$ we define $G_u = \frac{\displaystyle{\sum_{i\in U} \sum_{j \in U}} | \zeta_i - \zeta_j |}{2 \displaystyle{\sum_{i \in U} \sum_{j \in U} \zeta_j}}$
If $G_u = 0$ this means that the channel balance coefficients are equal.
In contrast, if $G_u = 1$ the channel balance coefficients are distributed in the most unequal way.

Note, that the Gini coefficient of a channel balance coefficients of a node $u$ takes the value $0$ if and only if its channel balance coefficients all take the same value.
This value will be exactly the same as the node's balance coefficient $\nu_u$.
Nodes which have channels with large differences in capacities will not be able to get a Gini coefficient of $0$ if we used absolute balances for the definition of the $\zeta$ values as smaller channels might be drained by larger ones during rebalancing operations.

Finally, $G$ denotes the imbalance of the network. $G = \displaystyle{\frac{1}{|V|}\sum_{v\in V}G_v}$. It is the mean of the imbalance values of all nodes in the network.
A perfectly balanced network would be achieved if $G$ takes the value of $0$ whereas the balance is poor if the value of $G$ is close to 1.

Our goal is to find a balance function $b$ which minimizes $G$ given a privacy aware payment channel network with initial distribution of funds $\tau_{u_1},\dots,\tau_{u_n}$.
The constraint to this optimization problem is that the total funds $\tau_u$ are fixed for every node $u \in V$ and any choice of the balance function $b$.
In a privacy aware payment channel network the distribution of funds $\tau_{u_1}\dots,\tau_{u_n}$ is not publicly known.\footnote{Even though the initial distribution could be guessed from the funding transactions in the case of the Bitcoin Lightning Network (c.f.: \url{https://github.com/lightningnetwork/lightning-rfc/issues/720}) for public channels it will change with the first payment that is not a rebalancing operation.}
In the same way the initial balance function is not publicly known.
As we lack knowledge about the global network state, we cannot apply standard optimization techniques such as gradient descent, conjugate gradient methods or simulated annealing.
We suggest to use an heuristic in which every participant executes some operations to improve its own balance which others support if it also improves their balance.

\subsection{Description of the rebalancing algorithm}
\label{sec:Algorithm}

The developed rebalancing algorithm uses a heuristic in which participants use the local knowledge and make local adjustments to estimate the optimization problem of finding $b$ such that $G$ is minimized. The algorithm therefore is privacy-aware. 
It executes as follows:
\begin{enumerate}
\item A node $u$ computes its node balance coefficient $\nu_u$.
\item $u$ then computes the channel balance coefficients $\zeta_{(u,v_1)},\dots,\zeta_{(u,v_d)}$ for its $d$ channels.
\item The node selects all channels $e=(u,v_i)$ for which its channel balance coefficient is higher 
than its node balance coefficient, i.e.~$C = \{(u,v_i) | \zeta_{(u,v_i)} - \nu_u\ > 0\}$\footnote{
  Note, that we do not need to take absolute values as $u$ will only be able to initiate a rebalancing operation by sending money which means decreasing its channel balance coefficient of $\zeta_{(u,v_i)}$ towards $\nu_u$.}.
\item From the candidate set $C$ a random channel $e=(u,v)$ is selected.
\item Now the node searches for a circular payment to itself along $e=(u,v)$ by choosing a path $p = [v,x_1,\dots,x_n,u]$. The amount of that payment should decrease the value of $\zeta_{(u,v)}$ to that of $\nu_u$ and can be computed as $a = c(e)\cdot (\zeta_{(u,v)}-\nu_u)$. The end of the circle should be a channel $(x_n,u)$ for which the channel balance coefficient $\zeta_{(u,x_n)}$ is smaller than the node balance coefficient $\nu_u$.\footnote{In one of the experiments we weaken this strong criteria and show empirically that it makes sense to drop it for the benefit of easier path finding at the cost of small oscillations of the algorithm.}
\item The node conducts the payment if all the nodes on the path $p$ agree to participate. It could happen that some nodes will only participate with a value smaller than $a$. As this is already progress $u$ will accept the suggested amount instead of being stubborn. 
\item Repeat all steps as long as the local balance coefficients are not even enough and as long as paths are found.
\end{enumerate}

When a circular path exists, it is easy to see that making a payment along this path will make the distribution of local balance coefficients for node $u$ and those of all participants more even as that was the condition under which they participated in the rebalancing operation.
If no such circular path can be found the local balance coefficients and thus the Network health stays constant.

\section{Experimental Setup}
\label{sec:setup}

We evaluated $4$ different strategies for finding rebalancing cycles. \texttt{cycles4} tests all cycles of length $4$ or smaller. In the same way \texttt{cycle5} tests all cycles of length $5$ or smaller. \texttt{foaf} tests most cycles of the friend of a friend network.\footnote{This approximation is achieved by using \texttt{cycle4} as a cycle base as all cycles of length 4 and smaller certainly are part of the friend of a friend network.}
\texttt{mpp} is short for multi-path payments and uses the same heuristic as \texttt{foaf} but will only take a $20^{th}$ of the maximum possible amount for rebalancing operation since it is supposed to split the rebalancing over various cycles.
Our approximation for \texttt{foaf} and \texttt{mpp} explains why the final results for these two strategies are not entirely monotonic but oscillating a bit.

We conducted our experiment on a snapshot of the public Lightning Network from October 2019.
That network view was retrieved from the gossip store of our lightning node which is accessible at \url{https://ln.rene-pickhardt.de}:
As channels are currently almost always opened by one side\footnote{Dual funded channels are not part of the protocol and no implementation is merged to any of the standard nodes} we randomly guessed who opened the channel by a coin flip and allocated the entire channel capacity to that node.
This had to be done as the gossip protocol does neither propagate any information about who is opening a channel nor who owns which funds.

\begin{figure}
 \centering
 \includegraphics[width=8cm]{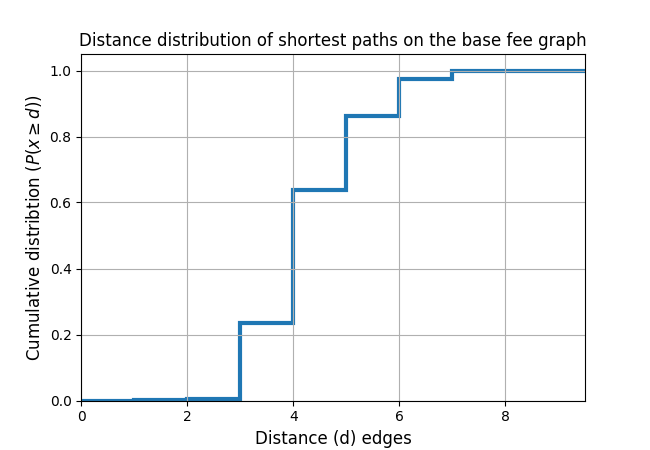}
 \caption{Probability mass of path-lengths of shortest paths between all pairs of nodes. As this is computed on the strongly connected component there will always be a path between each pair of nodes}
 \label{fig:cumulative_distance}
\end{figure}

After we randomly chose the founder of each channel we compute the largest strongly connected component to remove nodes which had the funds allocated in a way that would not allow any rebalancing.
The strongly connected component consists of $2707$ nodes and $24161$ edges.
The diameter of the strongly connected component had a value of 10 and the cumulative distance distribution of shortest paths between all pairs can be seen in~\cref{fig:cumulative_distance}.
Note that the notion of shortest paths with respect to fees depends on the amount that is being transferred as the fees on channels have a fixed base fee and a variable fee rate.
For the computation of shortest paths we have assumed a fixed payment size and, essentially, computed the shortest paths by only respecting the fixed base fee.
Shortest paths correspond to the cheapest paths.

Note that almost $40\%$ of shortest paths are longer than the cycles that our strategies have tested. 
This demonstrates that the single rebalancing operations which we conducted were not covering the entire network but only taking place within local parts of the network.
This is particularly true for the rebalancing in the friend of a friend network.

\begin{figure}
 \centering
 \includegraphics[width=8cm]{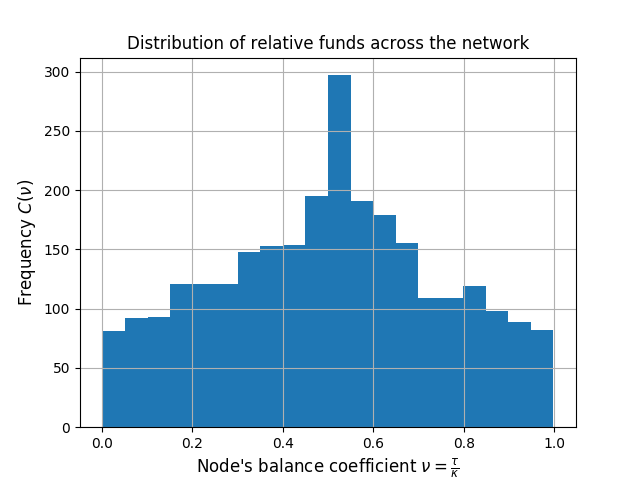}
 \caption{The histogram shows the distribution of relative funds $\nu$ for all nodes of the network.}
 \label{fig:initial_funds}
\end{figure}
From~\cref{fig:initial_funds} we can see the result of our random process of allocating funds to the nodes and selecting the largest strongly connected component.
It is plausible that $\nu=0.5$ is the mode of the histogram as the coin flip will statistically lead to as many channels with inbound capacity as channels with outbound capacity.

\begin{figure}
 \centering
 \includegraphics[width=8cm]{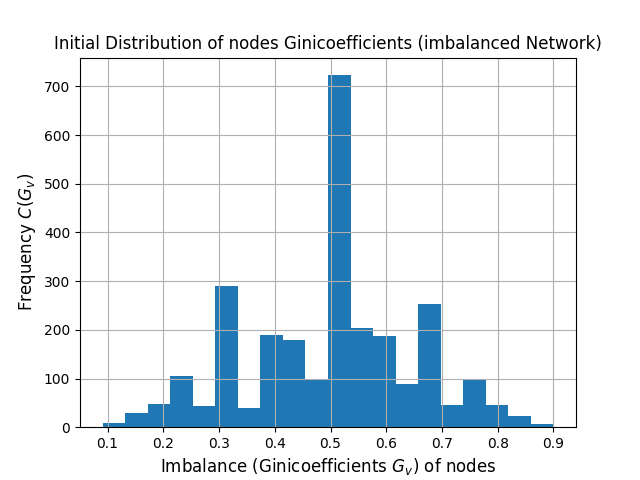}
 \caption{The distribution of Gini coefficients for each node over its distribution of channel balance coefficients $\zeta$ of the input network before any rebalancing experiment was taking place.}
 \label{fig:initial_ginis}
\end{figure}

\Cref{fig:initial_ginis} depicts the initial distribution of Gini coefficients for each node of the examined network.
This plot also reflects the random nature of the allocated funds already resulting in an average value as the total network imbalance very close to $0.5$.

In order to conduct the experiments we simulated the proposed algorithm and strategies in the following way.
For each channel we computed up to $5000$ rebalancing cycles following the defined strategy.
The number $5000$ was chosen to be as large as possible so that we were able to conduct the simulation within the $16$ GB of main memory of our machine.
In most cases there were less than $5000$ cycles available.
However sometimes we had to cut off the precomputed cycles as the hard cap of $5000$ was hit.

For each cycle we checked if a rebalancing was possible without having nodes on the path worsening their channel balance coefficients.
As this assumption is very strong the vast majority in rebalancing attempts along a cycle would fail or only work with an amount smaller than the one proposed by the initiator.

\section{Results}
\label{sec:results}

\begin{figure}
 \centering
 \includegraphics[width=7cm]{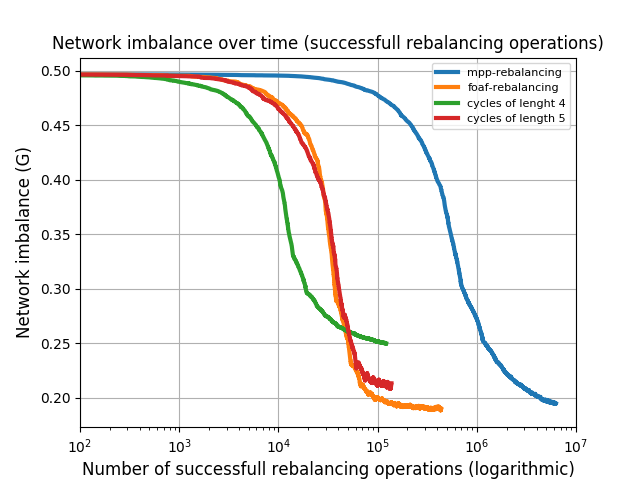}
 \caption{Comparing how the imbalance score of the network behaves with the amount of successfull rebalancing operations. As the rebalancing amounts are much smaller with multi path payments the x-axis has a logarithmic scale.}
 \label{fig:imbalancehovertime}
\end{figure}

\cref{fig:imbalancehovertime} confirms that the imbalance of the network is decreasing over time when running our simulation.
This means that the network is becoming more balanced.
We see that most improvement is happening quite fast with $10$k to $100$k rebalancing operations being necessary.
This is not more than $37$ rebalancing operations per node.
Taking into account that the average node degree is $9$ this means that on average each node needs to successfully execute about $4$ rebalancing operations per channel.  
This seams plausible in the sense that every channel gets rebalanced at least once.\footnote{Note, that due to the collaborative behaviour of nodes a node might also get its channels rebalanced in the rebalancing attempts of other nodes.}
The multi-path rebalancing needs more operations which makes sense as we only rebalanced for a $20^{th}$ of the possible amount in the multi-path case since we wanted to have multiple other cycles to rebalance that particular channel.
While \texttt{cycle4} seems to improve the balance more quickly it does not reach the minimum imbalance as well as the other strategies.
In particular the algorithm converges to a local minimum rather quickly.
Our experimental runs resulted in states that are not identical when we shuffled the order of cycles that we probed, however, the differences were negligible. 
We guess that the local minimum might be close to global minimum of the optimization problem.

\begin{figure}
 \centering
 \includegraphics[width=8cm]{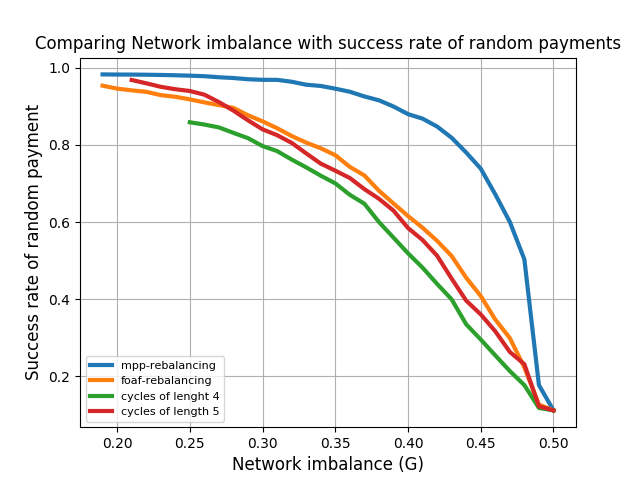}
 \caption{Every time the imbalance score reached a new low to a two digit decimal we computed the probability for a random payment of the smallest possible amount to succeed on the cheapest path.}
 \label{fig:imba_vs_success}
\end{figure}

In~\cref{fig:imba_vs_success} we see that a lower imbalance score $G$ yields a higher success rate of payments as one would expect.
The success rate is measured by attempting a payment of $1$ Satoshi between all pairs of nodes and counting the relative amount of payment attempts for which the cheapest path from the source to the destination was successful.
We computed the success rate of every time the balance score hits a new score rounded to two decimals.
It is the first experiment that indicates that the defined imbalance score was well chosen. 
Again the multi-path payment strategy sticks out as it achieves a rather high success rate while the network is still rather imbalanced.
A success rate of $80\%$ is reached for an imbalance score of $0.43$.
Comparing this to~\cref{fig:imbalancehovertime} we see that about $250$k rebalance operations are necessary to achieve this success rate.
Even the worst performing algorithm achieves this success rate with far less than $100$k rebalancing operations and a much better overall imbalance score. 

This effect - while expected as multi-path rebalancing uses smaller amounts - becomes more visible in the next experiment.
As $1$ Satoshi payments are not too useful in a real world setting we checked if the overall amounts that can be routed increase in a better balanced network.

\begin{figure}
 \centering
 \includegraphics[width=8cm]{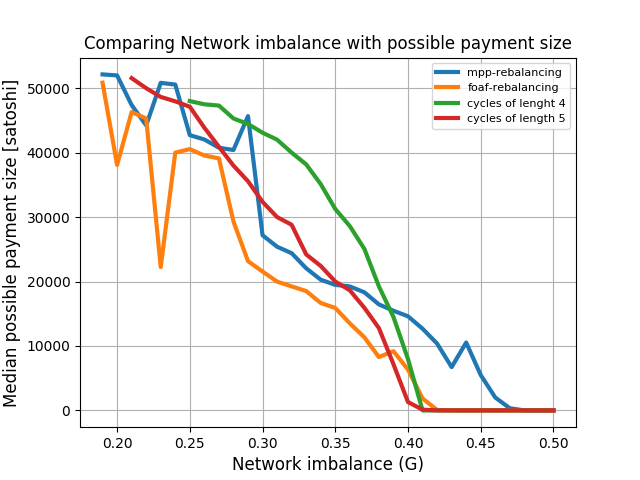}
 \caption{The imbalance of the network is plotted against the median value of payments that can be fulfilled between all pairs of nodes on the first attempt along the cheapest path. Note that failed payments are part of the statistics and counted as payments which can send $0$ Satoshi.}
 \label{fig:imba_vs_payment_size}
\end{figure}

In~\cref{fig:imba_vs_payment_size} we compare the imbalance of the network with the median possible payment amount which is computed as follows:
For all pairs of shortest paths on the base fee graph we look at the amount that could be forwarded along that path.
Finally we take the median of those values.
This means that at least $50\%$ of payment pairs are able to forward this amount.
Again we see that the lower the imbalance score becomes the higher the median possible payment size gets.
This suggests that statistically a more balanced network is able to successfully route higher payment amounts on the first try.
All four strategies are achieving a median possible payment amount of roughly $50000$ Satoshi.
This means that in a balanced network $50\%$ of payment pairs are able to successfully conduct a payment of $0.5$ mBTC along the cheapest path. 
The result also indicates that the strategy for probing rebalancing cycles seems to have little to no impact to the final abilities of the network to perform payments.


\begin{figure}
 \centering
 \includegraphics[width=8cm]{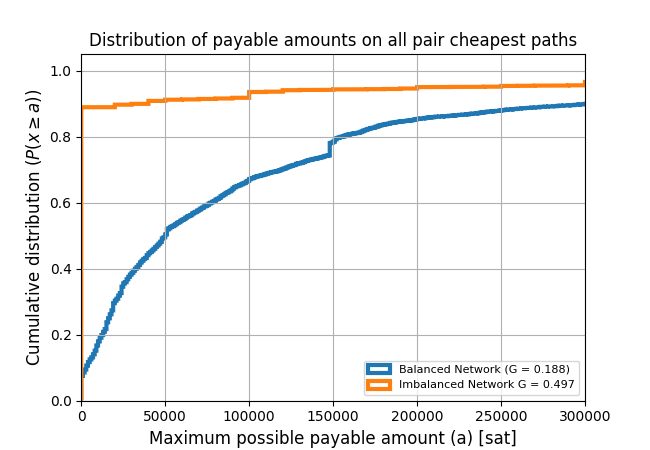}
 \caption{The maximal payable amounts on all pairs of cheapest paths on the initial imbalanced network and on the network after applying the friend of a friend rebalancing strategy.}
 \label{fig:cdf_paymentsize}
\end{figure}

We look a little bit closer at the data point of the most balanced network with the friend of a friend strategy and compare it to the imbalanced network.
Instead of just looking at the median we study the cumulative distribution function of the histogram in~\cref{fig:cdf_paymentsize}.
We can see that in the imbalanced network almost $88.8\%$ of the paths are not able to forward a single Satoshi (meaning a $11.2\%$ success rate).
This number drops below $1.7\%$ (meaning a $98.3\%$ success rate) for the balanced network.

\begin{figure}
 \centering
 \includegraphics[width=8cm]{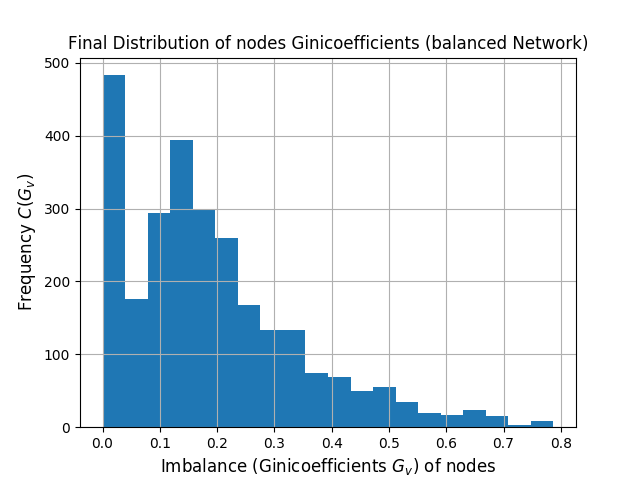}
 \caption{The distribution of Gini coefficients for each node over its distribution of channel balance coefficients $\zeta$ of the network after applying the foaf rebalancing strategy.}
 \label{fig:final_gini}
\end{figure}

For the friend of a friend network we have also plotted the final distribution of Gini coefficients in~\cref{fig:final_gini}. 

\begin{figure}
 \centering
 \includegraphics[width=8cm]{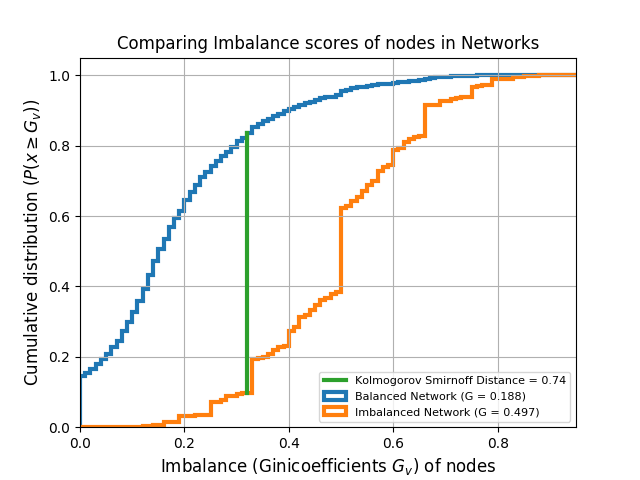}
 \caption{Comparing the distribution of the node's Gini coefficients before and after rebalancing with the friend of a friend strategy.}
 \label{fig:cdf_gini}
 \end{figure}

For a better comparison with~\cref{fig:initial_ginis} we provide the cumulative distribution function of both distributions with a smaller bin size in~\cref{fig:cdf_gini}.

The Kolmogorov-Smirnoff distance takes a value of $0.74$.
This verifies the statistical significance of the rebalancing heuristic in comparison to staying with the imbalanced network.
The median value of the Gini coefficients in the best balanced network we could acquire is about $0.15$.
In comparison on the imbalanced network the median Gini coefficient takes the value of $0.5$.

Finally, we tracked the routing fees that nodes would have to pay for their rebalancing operations and that they would earn while participating in the rebalancing operation of others.
 \begin{figure}
 \centering
 \includegraphics[width=8cm]{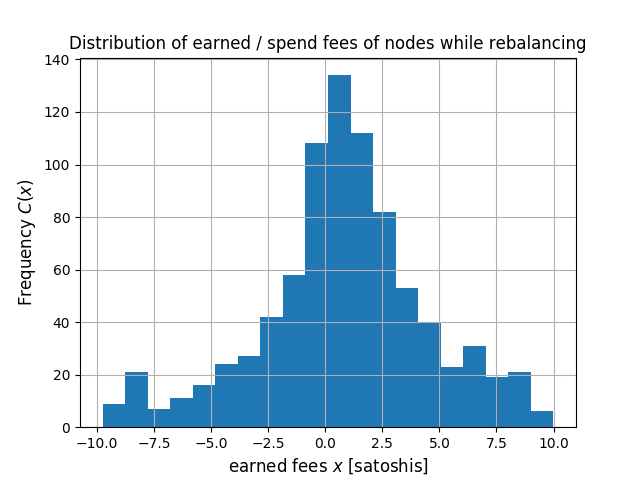}
 \caption{Total fees earned by nodes during all rebalancing operations. Negative fees mean that the node paid more fees than it earned.}
 \label{fig:fees}
 \end{figure}

\Cref{fig:fees} shows that the overall fees are normally distributed around $0$.
This means that the expected value of fees that need to be paid or are being earned if all nodes participate in proactive rebalancing is $0$.
An expected value of $0$ is a strong indicator that fees should not play a role for rebalancing operations if those are beneficial for all nodes that are asked to take part in the rebalancing operation.
We zoomed heavily in so that only $50\%$ of the nodes have been depicted in this view due to the large variance of this distribution.

\section{Discussion}
\label{sec:conclusion}

We empirically verified that the overall ability of the network participants to route payments on random paths is improved a lot in a balanced network. 
As privacy aware payment channel networks use source-based path finding without knowing the constraints of the channels along an attempted path, the likelihood of conducting a successful payment with few attempts becomes considerably higher in a balanced network.
Our results show that nodes do not need to strive for $\nu_u = 0.5$ and perfect local balance of their channels in order to produce high success rate of random payments.
It is sufficient for the overall network to be balanced if every node tries to optimize their channel balance coefficients be almost equal. 
This can be achieved by making circular rebalancing payments.

We have proposed a greedy rebalancing heuristic which we have tested with $4$ different strategies to select rebalancing cycles.
We could see that the results of these four strategies do not significantly differ for the overall performance of the network.
While multi-path rebalancing achieves the actual best values on both success rates and median payment size the lead is negligible and stands in no ratio to the extra amount of rebalancing operations that need to take place.

Computing the rebalancing cycles is the most expensive operation in our simulation since we had to do it for every node of the network.
If the greedy heuristic is implemented in a real payment channel network this is not too expensive for a single node.
This holds in particular true if those cycles will only be searched within the friend of a friend network which is typically small enough.
As the results of the different strategies for finding cycles did not vary and for the previous reason we propose to stick to rebalancing in the friend of a friend network.
In a real payment channel network this computation would only have to be conducted 
if the distribution of channel balance coefficients becomes too uneven,
which is easily monitored after every payment or routing attempt without additional performance overhead.

Finally, we empirically showed that the earned and paid fees for rebalancing operations are normally distributed around $0$ Satoshi suggesting that this is a zero sum game if every node participates in proactive rebalancing.
Routing fees might discourage nodes from proactively rebalancing as they might hope to earn fees if they wait for others to rebalance their channels.
This behaviour is referred to as the tragedy of the commons\footnote{c.f.~\url{https://en.wikipedia.org/wiki/Tragedy_of_the_commons}} and would be mitigated with a fee free rebalancing protocol.
As fees are empirically a zero sum game and to incentivize the participation of the rebalancing it seems reasonable to omit fees for rebalancing operations if those are beneficial for every participant as in the experiment.\footnote{This would of course need a protocol change in the lightning network protocol as the current onion routing could not guarantee to the routing noes that they process a fee free rebalancing payment.}

In a similar way as Internet routers collaboratively solve the path finding problem by exchanging routing tables, we furthermore suggest for the network to work collaboratively towards achieving a good balance by locally sharing rebalancing hints. 
As balances of channels in the friend of a friend network can easily be probed, it seems plausible in privacy aware payment channel networks to include a communication protocol with which a node signals to their partners that it wants to rebalance a certain channel with a certain amount.
Instead of a certain amount a node could just signal if it needs inbound or outbound liquidity on a channel. 
We think that our suggestion for sharing channels that need a rebalancing operation does not reveal more information as can currently be collected with probing attacks.
However, sharing rebalancing hints gives out perfect information to compute available rebalancing cycles.

The rebalancing is particularly important as privacy aware payment channel networks use source based onion routing and technically have to guess a path for making a payment.
When those initially guessed paths provide a high success rate with high payment amounts there are considerable benefits for the overall functioning of the payment channel network. 
The user experience is improved, the payment delays are lowered, and the amount of messaging and coordination is minimised.

\section{Future Work}
\label{sec:future}

As the simulation was already consuming a lot of computational resources we did not check if the algorithm greedy heuristic is stable under concurrent payments and rebalancing operations taking place.
In a similar gist simulations on larger networks would be interesting but require more significant computational resources and time to be conducted.

The current experiments have assumed a stable network topology. 
However, every channel opening or closure, as well as every single payment, will change the imbalance of the network.
It is therefore necessary to study how it adapts if the topology changes in real time due to opening or closing channels as well as a reallocation of funds due to payments that are taking place.
It would be in particular interesting to see how this heuristic works in combination with JIT-Routing as this would unbalance nodes to provide liquidity which our heuristic would then redistribute over the friend of a friend network.
This would occur along all nodes on a path in intersecting friend of a friend networks.
It would be interesting to see how this algorithm behaves together with Atomic Multipath Payments which are in the style of~\cite{bagaria2019boomerang} redundant but along our results carry at most $50000$ Satoshi. 

Another extension of the work is to check how the heuristic performs if only a fraction of nodes participate in the sharing of rebalancing hints and fee free rebalancing protocol.

Current implementations of the Lightning Network provide autopilots to the users which help the users to automatically open channels.
The results from this research should be used in those autopilots and an investigation how exactly this might be useful would be of high interest.

The balance metric used could be used as a measure of network health in the context of random path payments. 
However, random path payments are only one of the possible scenarios of payments flowing in a payment network. 
Thus, a natural extension of this work is to 
study the relationship between the proposed balance and network health. Can nodes define their local health in various ways and what would happen with the greedy algorithm if different health definitions are followed by individual nodes?
For example routing nodes or liquidity providers might very well want to have high values of $\zeta$ on some of their channels where as they might be willing to accept low values for $\zeta$ on other channels.
In a similar way merchants will probably prefer many channels with low values of $\zeta$.
The proposed algorithm could potentially be exchanged to an updated version in which nodes decide their own health metric,
instead of using balance as the measure of network health.

\section{Acknowledgements}
\label{sec:ack}
We are grateful to Vivek Bagaria and Joachim Neu for an interesting discussion about their and our work that initiated this particular research. We thank Stefan Richter for helpful comments on early drafts of this work. A special thank goes to Thomas Gottron, Christian Decker and Dmytro Piatkivskyi. Finally we thank the donors to \url{https://tallyco.in/s/lnbook} and Patreons supporting open knowledge. 

\bibliography{lightningNetworkHealth}
\bibliographystyle{plain}

\end {document}